\documentclass[11pt,draftcls,onecolumn]{IEEEtran}
\usepackage{amsfonts}
\usepackage{amsmath}
\usepackage{amssymb}
\usepackage{latexsym}
\usepackage{epsfig}
\usepackage{pstricks}
\usepackage{psfrag}
\usepackage{bm}
\usepackage{pst-grad}
\usepackage{graphicx}
\usepackage{subfigure}
\usepackage[nosort]{cite}

\newenvironment{proof}[1][Proof]{\begin{trivlist}
\item[\hskip \labelsep {\bfseries #1}]}{\end{trivlist}}

\newcommand{\qedclosed}{$\blacksquare$}

\newtheorem{theorem}{Theorem}

\newtheorem{lemma}{Lemma}
\newtheorem{corollary}{Corollary}

\newtheorem{example}{Example}

\newenvironment{remark}{\textit{Remark:}}

\newcommand{\cov}{\textrm{cov}}

\newcommand{\tr}{\textrm{tr}}

\newcommand{\R}{\mathbb{R}}

\newcommand{\thetam}{\theta_{\min}}

\newcommand{\btheta}{\bm{\theta}}
\newcommand{\es}{\bm{\hat{s}}}
\newcommand{\bs}{\bm{s}}
\newcommand{\bPhi}{\bm{\Phi}}
\newcommand{\bepsilon}{\bm{\epsilon}}
\newcommand{\by}{\bm{y}}
\newcommand{\bdelta}{\bm{\delta}}
\newcommand{\bx}{\bm{x}}
\newcommand{\bDelta}{\bm{\Delta}}
\renewcommand{\S}[1]{\mathcal{S} \{\bm{\Phi}_{#1}\} }

\providecommand{\calN}{\mathcal{N}}

\providecommand{\fi}{\dft{i}}

\begin{document}

% paper title
\title{An Estimation Theoretic Approach for Sparsity Pattern Recovery in the Noisy Setting}

\author{Ali Hormati, \IEEEmembership{Student Member, IEEE}, Amin Karbasi, \IEEEmembership{Student Member, IEEE},\\ Soheil Mohajer, \IEEEmembership{Student Member, IEEE}, and Martin Vetterli, \IEEEmembership{Fellow, IEEE}%
\thanks{The authors are with the School of Computer and Communication
Sciences, Ecole Polytechnique F\'{e}d\'{e}rale de Lausanne (EPFL), CH-1015
Lausanne, Switzerland (e-mails: \{ali.hormati, amin.karbasi, soheil.mohajer,
martin.vetterli\}@epfl.ch). Martin Vetterli is also with the Department of
Electrical Engineering and Computer Sciences, University of
California, Berkeley, CA 94720, USA.}%
\thanks{This work was supported
by the Swiss National Science Foundation under grants
NCCR-MICS-51NF40-111400 and 200020-103729. The material in this work was presented in part at the IEEE International Symposium on Information Theory, Seoul, Korea, June 2009.}}

\maketitle

% Abstract
\begin{abstract}
Compressed sensing deals with the reconstruction of sparse signals using a small number of linear measurements. One of the main challenges in compressed sensing is to find the support of a sparse signal. In the literature, several bounds on the scaling law of the number of measurements for successful support recovery have been derived where the main focus is on random Gaussian measurement matrices.

In this paper, we investigate the noisy support recovery problem from an estimation theoretic point of view, where no specific assumption is made on the underlying measurement matrix. The linear measurements are perturbed by additive white Gaussian noise. We define the output of a support estimator to be a set of position values in increasing order. We set the error between the true and estimated supports as the $\ell_2$-norm of their difference. On the one hand, this choice allows us to use the machinery behind the $\ell_2$-norm error metric and on the other hand, converts the support recovery into a more intuitive and geometrical problem. First, by using the Hammersley-Chapman-Robbins (HCR) bound, we derive a fundamental lower bound on the performance of any \emph{unbiased} estimator of the support set. This lower bound provides us with necessary conditions on the number of measurements for reliable $\ell_2$-norm support recovery, which we specifically evaluate for uniform Gaussian measurement matrices. Then, we analyze the maximum likelihood estimator and derive conditions under which the HCR bound is achievable. This leads us to the number of measurements for the optimum decoder which is sufficient for reliable $\ell_2$-norm support recovery and shows that the performance of the optimum decoder has only a 9 dB gap compared to the HCR lower bound. Using this framework, we specifically evaluate sufficient conditions on the number of measurements for uniform Gaussian measurement matrices.\nocite{KarbasiHMV09}
\end{abstract} 

% Keywords

% Introduction
\section{Introduction}
Linear sampling of sparse signals, with the number of measurements close to their sparsity level, has recently
received a lot of attention under the names of compressed sensing (CS), compressive
sampling  or sparse sampling~\cite{Donoho06,CandesRT06, CandesT06, BluDVMC08}. A $k$-sparse signal $\bm{\theta} \in \mathbb{R}^p$ is defined as a signal with $k\!\ll\! p$ nonzero expansion coefficients in some orthonormal basis
or frame. The goal of compressed sensing is to find measurement matrices $\bm{\Phi}_{m\times p}$,
followed by reconstruction algorithms which allow robust recovery
of sparse signals using the least number of measurements $m$, and low
computational complexity; see for example~\cite{TroppG07, CandesT05, DonohoTDS, SarvothamBB06, CormodeM06, Tropp04}.

Support recovery refers to the problem of estimating the positions of the non-zero entries of $\btheta$, based on a set of observations. In the noiseless setting, the optimal algorithm requires $m = k + 1$ samples at the expense of high computational complexity to obtain the true support set~\cite{FengB96} while $m = \mathcal{O}(k \log{(p/k)})$ measurements are needed for the reconstruction algorithms based on linear programming~\cite{ChenDS98}. In the same context, it is shown that $m=2k+1$ samples are sufficient for shift-invariant measurement matrices using recovery algorithms based on annihilating filters~\cite{VetterliMB02}.

In practice, however, all the measurements are noisy due to physical restrictions, quantization precision, etc. A large body of recent work has established bounds on the number of measurements required for successful support set recovery in the noisy setting. Denoting $\theta_{\text{min}}$ as the minimum non-zero coefficient of the sparse vector $\btheta$, the authors in~\cite{Wainright06, MeinshausenB06} derived the scaling law on the number of measurements as a function of $(p,k,\theta_{\text{min}})$ for the $\ell_1$-constrained quadratic programming, also referred to as  Lasso, to recover the sparsity pattern. In the context of the optimal decoding algorithm, the results in~\cite{Wainright07,FletcherRG08} provide necessary and sufficient conditions for the perfect support recovery under the Gaussian measurement ensemble. Considering a fractional support recovery, the study in~\cite{ReevesG08} provides a set of necessary and sufficient conditions on the required number of measurements as a function of the fraction of the support that can be reliably recovered.

In this paper, we look at the support recovery problem from an estimation theoretic point of view, where the error metric between the true and the estimated support is the $\ell_2$-norm of their difference. In some applications, e.g.~\cite{CotterR02}, it is important that the recovered sparsity pattern be as close as possible to the true support set. In these cases, the $\ell_2$-norm error metric comes as an appropriate option where the assigned penalty is quadratically proportional to the distance. Moreover, this choice allows us to use the machinery behind the $\ell_2$-norm error metric, which makes the theorems and the proofs geometrical and more intuitive. While no specific assumption is made on the underlying measurement matrix, we assume that the linear measurements are perturbed by additive white Gaussian noise. Since the positions of the nonzero entries of $\bm{\theta}$ forms a set of $k$ discrete values (e.g., integers between  $1$ and $p$), the support recovery problem can be regarded as estimating restricted parameters. This leads us to use the Hammersley-Chapman-Robbins (HCR) bound which provides a lower bound on the variance of any unbiased estimator of a set of restricted parameters \cite{Hammers50,Chapman51}. The HCR bound is a generalization of the Cramer-Rao (CR) bound~\cite{Stoica00} and holds under much weaker regularity conditions, while giving substantially tighter bounds in general. Using the HCR bound, we specifically derive in a straightforward manner the necessary conditions on the required number of measurements for the standard Gaussian ensemble.

Of equal interest are the conditions under which the HCR bound is achievable (tight). To this end, we study the performance of the maximum likelihood estimator (MLE) and derive conditions under which it becomes unbiased and achieves the HCR bound. In particular, this leads us to the sufficient conditions on the number of measurements for reliable $\ell_2$-norm support recovery using the standard Gaussian measurement ensemble. Note that when the error of the $\ell_2$-norm support recovery vanishes, so does that of a regular support recovery problem with the $\{0,1\}$ error metric. Therefore, the derived sufficient condition also applies to the $\{0,1\}$ error metric support recovery.

The organization of the paper is as follows. In Section~\ref{sec:problem}, we provide a more precise formulation of the problem. We derive the HCR bound for the support recovery problem in Section~\ref{sec:hcr} which is followed by deriving necessary conditions on the number of measurements for the standard Gaussian measurement ensemble. By studying the performance of the MLE in Section~\ref{sec:ml}, we derive conditions under which the HCR bound becomes achievable. Finally, under the standard Gaussian measurement ensemble, we identify the sufficient number of measurements for reliable $\ell_2$-norm support recovery. 

% Previous Work
\section{Previous Work}
The problem of sparsity recovery has received considerable attention in the literature in both the noiseless and noisy settings, see e.g.,~\cite{CandesT05, Wainright06, Wainright07, FletcherRG08, ReevesG08, DonohoET06, AkcakayaT08}. The results focus on the asymptotic scaling of the number of measurements for almost-sure success of the reconstruction of sparse inputs. In this section, we give an overview of the previous work which is more related to the results of this paper.

The work in~\cite{Wainright07} provides necessary and sufficient conditions on the number of measurements in the high-dimensional and noisy setting for reliable sparsity recovery using an optimal decoder. In that setup, the measurements are contaminated by i.i.d. Gaussian noise and the analysis is high dimensional, meaning that the sparsity level $k$, the signal dimension $p$ and the number of measurements $m$ tend to infinity simultaneously. Under the condition $(m-k)\:\theta_{\text{min}}^2 \!\rightarrow \!+\infty$, the author derives the following sufficient condition for asymptotic reliable recovery of the optimal decoder
 \begin{equation}
 \label{WainrightSuff}
   m > C \: \max\left\{k\log{(p/k)},\: \frac{1}{\theta_{\text{min}}^2} \log{(p-k)}\right\},
 \end{equation}
  where $C > 0$ is a fixed constant. Moreover, it is also shown in~\cite{Wainright07} that
  \begin{equation*}
    m >  \frac{C'}{\theta_{\text{min}}^2} \log\frac{p}{k},
  \end{equation*}
  is a necessary condition for some fixed constant $C' > 0$. By simplifying the sufficient condition~\eqref{WainrightSuff} in the sublinear sparsity regime $k = o(p)$, it is shown that the number of measurements required by the $\ell_1$ constrained quadratic programming (Lasso) given by $m = \Theta(k\log{(p-k)})$~\cite{Wainright06} achieves the information-theoretic necessary bound.

In~\cite{FletcherRG08}, the authors derive the necessary scaling
\begin{equation}
\label{equ:Goyalnecessary}
  m > \frac{2}{\mathsf{MAR}\cdot\mathsf{SNR}}\:k\log{(p-k)}+k-1,
\end{equation}
for uniform i.i.d. Gaussian measurement ensemble which is true at any finite $\mathsf{SNR}$ and for all algorithms. The term $\mathsf{MAR}$ indicates the minimum-to-average ratio of the input sparse signal. Moreover, they show that for a fixed $\mathsf{SNR}$ and $\mathsf{MAR}$,  the simple maximum correlation estimator (MCE) achieves the same scaling as in~\eqref{equ:Goyalnecessary}. The MCE selects the indices of the $k$ columns of the measurement matrix having the highest correlation with the measurement vector. More precisely, the results indicate that MCE needs
\begin{equation}
\label{equ:GoyalMCSufficient}
  \frac{8(1+\mathsf{SNR})}{\mathsf{MAR}\cdot\mathsf{SNR}}\:k\log{(p-k)}
\end{equation}
measurements to succeed with high probability. Therefore, the simple MCE also achieves the same scaling law as Lasso.

In a more general setting, the support recovery with some distortion measure has been considered in~\cite{SarvothamBB06, ReevesG08, FletcherRG07}. The results in~\cite{ReevesG08} show that if the $\mathsf{SNR}$ does not increase with the signal dimension, the exact support recovery is not possible. Moreover, they show that partial support recovery is possible with a bounded $\mathsf{SNR}$ per sample which indicates that a finite rate per sample is sufficient. In this regard, our work can be viewed as the support recovery problem with the $\ell_2$-norm distortion measure. In the following, we explain our setup for the estimation theoretic approach of support recovery.

% Problem Statement
\section{Problem statement}
\label{sec:problem}
In this paper, we consider a deterministic signal model, in which $\btheta\in\R^p$ is a fixed but unknown vector with \emph{exactly} $k$ non-zero entries. We refer to $k$ as the signal sparsity, $p$ as the signal dimension, and define the support vector $\bs(\btheta)$ as the positions of the non-zero elements of $\btheta$. More precisely,
\begin{equation*}
\bs(\btheta)\triangleq(n_1,n_2,\dots,n_k),
\end{equation*}
where we assume that $n_1<n_2<\dots<n_k$. The corresponding non-zero entries of $\btheta$ form a vector
\begin{equation*}
\btheta_s\triangleq(\theta_{n_1},\theta_{n_2},\dots,\theta_{n_k}).
\end{equation*}
Suppose we are given a vector of $m$ noisy observations $\by\in\R^m$ of the form
\begin{equation*}
\by=\bPhi\btheta+\bepsilon,
\end{equation*}
where $\bPhi\in\R^{m\times p}$ is the measurement matrix and $\bepsilon\sim \calN\left(0,\sigma^2 \mathbf{I}_{m\times m}\right)$ is additive i.i.d. Gaussian noise. Throughout this paper, we assume that $\sigma^2$ is fixed; since any scaling of $\sigma^2$ can be accounted for in the scaling of $\btheta$. Let $\bx =\bPhi\btheta$, $\bPhi_{\bs}$ denote the matrix composed of the columns of $\bPhi$ at positions indexed by $\bs(\btheta)$, and $\S{\bs}$ denote the column span  of $\bPhi_{\bs}$. Since there are $N={p \choose k}$ subspaces of dimension $k$, a number from $1$ to $N$ can be assigned to them and w.l.o.g., we assume that $\bx$ belongs to the first subspace $\S{\bs_1}$. From now on, for simplicity we refer to the first subspace as $\S{\bs}$. Moreover, we need to assume that any $2k$ columns of the measurement matrix $\bPhi$ are linearly independent. Under this assumption, we have $\btheta \neq \btheta' \Leftrightarrow \bx \neq \bx'$, i.e., there is a one-to-one correspondence between $k$ sparse vectors $\btheta$ and their images $\bx$.

Due to the presence of noise, $\btheta$ cannot be recovered exactly. However, a sparse-recovery algorithm outputs an estimate $\btheta'$. In the support recovery problem, we are only interested in estimating the support. To that end, we can consider different performance metrics for the quality of estimation. In~\cite{Wainright06}, the measure of error between the estimate and the true signal is a $\{0,1\}-$valued loss function,
\begin{equation*}\label{metric1}
\rho_1(\bs,\bs')\triangleq\mathbb{I}\left( \bs\neq\bs'\right),
%\rho(\theta,\theta'):=\mathbb{I}\left(\{\hat{\theta_i}\neq 0,\forall i\in S\}\cap \{\hat{\theta_i}=0,\forall i\notin S\} \right).
\end{equation*}
where $\mathbb{I}(\cdot)$ is the indicator function. This metric is appropriate for the exact support recovery. In this work, we are interested in an approximate support recovery where the goal is to recover a sparsity pattern as close as possible to the true support set. For this purpose, we consider the following $\ell_2$-norm error metric
\begin{equation*}\label{metric2}
\rho_2(\bs,\bs')\triangleq\|\bs-\bs'\|^2,
\end{equation*}
where throughout this paper, $\|\cdot\|$ refers to the Euclidean norm.
Note that $\rho_2(\bs, \bs')=0  \Leftrightarrow \rho_1(\bs,\bs')=0$.

As is mentioned in~\cite{Wainright07}, $\mathsf{SNR}$ alone is not a suitable quantity for the support recovery problem. It is possible to generate a set of problem instances for which the support recovery becomes arbitrarily unreliable, in particular, by letting the smallest coefficient go to zero (assuming that $k>1$) at an arbitrarily rate, even though the $\mathsf{SNR}$ becomes arbitrarily large by increasing the rest. As he also observed, the magnitude of the smallest nonzero entry
of $\btheta$ is prominent in the phrasing of results. Hence, we define
\begin{equation*}
\theta_{\min}=\min_{i\in \bs}|\theta_i|.
\end{equation*}
 In particular, our results apply to any unbiased estimator that operates over the signal class
\begin{equation*}
C(\theta_{\min})=\{\btheta\in\R^p:|\theta_i|\geq\theta_{\min}\ \forall i\in \bs(\btheta)\}.
\end{equation*}
Our analysis is high dimensional in nature, in the sense that the signal dimension $p$ goes to infinity. More precisely, we say the $\ell_2$-norm support recovery is reliable if
\begin{equation}\label{asym}
\lim_{p\rightarrow \infty}\rho_2(\bs(\btheta),\es(\btheta))=0,
\end{equation}
for any $\btheta\in C(\theta_{\min}) $ under some scaling of $\{\theta_{\min}, k, m\}$ as a function of $p$, where $\es(\btheta)$ is the estimated support of $\btheta$. For unbiased estimators, (\ref{asym}) is equivalent to
\begin{equation*}
\lim_{p\rightarrow \infty}\tr\:[\cov(\es(\btheta))]=0,
\end{equation*}
where
\begin{equation*}
\cov(\es(\btheta)) = \mathbb{E}\left[ (\es(\btheta)-\mathbb{E}[\es(\btheta)])^T(\es(\btheta)-\mathbb{E}[\es(\btheta)])\right],
\end{equation*}
and $\tr[\cdot]$ is the matrix trace operation. Since the support estimation is based on $\by$, with a slight abuse of notation, we also denote it by $\es(\by)$.

With this setup, our first goal is to find necessary conditions on parameters $\{p, m, k, \thetam\}$ which should be satisfied by any unbiased estimator for reliable $\ell_2$-norm support recovery. The results are applicable to any measurement matrix and we specifically evaluate it for the standard Gaussian measurement matrices. Our second goal is to find sufficient conditions for the successful support recovery using the optimum decoder. We show that under appropriate conditions, the performance of the optimum decoder is close to the theoretical lower bound for the performance of the unbiased support estimators. Again, as a special case, we evaluate the sufficient conditions for standard Gaussian measurement matrices.

% HCR bound
\section{Hammersley-Chapman-Robbins Bound }
\label{sec:hcr}
The Cramer-Rao (CR) bound is a well-known tool in statistics which provides a lower bound on the variance of the error of any unbiased estimator of an unknown deterministic parameter $\delta$ from a set of measurements $\by$~\cite{Stoica00}. More specifically, in a single parameter scenario, the estimated value $\hat{\delta}$ satisfies
\begin{equation}
\label{Cramer}
\mbox{var}(\hat{\delta}) \geq \frac{1}{- \int_{\R} \frac{\partial^2\ln{\mathbb{P}(y;\delta)}}{\partial \delta^2}\mathbb{P}(y;\delta)dy},
\end{equation}
where $\mathbb{P}(y;\delta)$ is the pdf of the measurements which depends on the parameter $\delta$. As~\eqref{Cramer} suggests, the CR bound is derived for estimating a continuous parameter.

In many cases, there is \textsl{a priori} information on the estimated parameter which restricts it to take values from a predetermined set. An example is the estimation of the mean of a normal distribution when one knows that the true mean is an integer (see the example below). In such scenarios, the Hammersley-Chapman-Robbins (HCR) bound provides a stronger lower bound on the variance of any unbiased estimator~\cite{Hammers50,Chapman51}. More precisely, let us  assume that the set of observations $\by=(y_1,y_2,\dots,y_m)$ are drawn according to a probability distribution with density function $\mathbb{P}(\by;\bdelta)$ where $\bdelta$ is a parameter belonging to some parameter set $\bDelta$ (e.g., the set of integer numbers) and completely characterizes the pdf. In addition, the sequence $\bdelta$ is partitioned into two subsequences $\bdelta=(\bdelta_1,\bdelta_2)$, where we are only interested in estimating the parameters included in the subsequence $\bdelta_1$. Let  $\bm{\hat{\delta}}_1(\by)$ denote an unbiased estimator of $\bdelta_1$. Given the above definitions, we recall the following result.

\begin{theorem}[\hspace{-0.01cm}\emph{\cite{Hammers50,Chapman51}}]
The trace of the covariance matrix of any unbiased estimator of $\bdelta_1$ is bounded below by
\begin{equation}
\label{Hammersley}
\mbox{tr}[\mbox{cov}(\bm{\hat{\delta}_1})] \geq \sup_{ \bdelta' \neq \bdelta}\frac{\|\bdelta_1-\bdelta'_1\|^2}{ \int_{\mathbb{R}^m} \frac{\mathbb{P}^2(\by;\bdelta')}{\mathbb{P}(\by;\bdelta)}d\by-1},
\end{equation}
in which $\bdelta'=(\bdelta'_1,\bdelta'_2) \in \bDelta$. The set $\bDelta$ is chosen so that $\bdelta'$ takes values according to the \textsl{a priori} information.
\end{theorem}

\begin{example}
For clarity, let us consider the performance of any unbiased estimator of (only) the mean of a normal distribution based on independent samples of size $m$, i.e., $\by=(y_1,y_2,\dots,y_m)$. In this case, $\bdelta=(\mu, \sigma^2)$, $\delta_1=\mu$, $\delta_2= \sigma^2$ and
\begin{equation*}
\mathbb{P}(\by;\bdelta)=(\frac{1}{\sqrt{2\pi \sigma^2}})^m e^{-\frac{1}{2\sigma^2}\sum_{i=1}^m(y_i-\mu)^2}.
\end{equation*}
Let $\hat{\mu}(\by)$ denote an unbiased estimator of $\mu$ which is the parameter we want to estimate. When there is no prior information on $\mu$, it follows from the CR bound that
\begin{equation}
\label{cr-gaussian}
\mbox{var}(\hat{\mu})\geq \sigma^2/m.
\end{equation}
Once the mean is restricted to be an integer, we may write $\delta_1=\mu$ and  $\delta'_1=\mu+\alpha$, where $\alpha$ is a non-zero integer. Then, upon integration in~\eqref{Hammersley} we get
\begin{eqnarray}
\label{hcr-gaussian}
\mbox{var}(\hat{\mu})&\geq& \max_{\alpha\neq 0}\frac{\alpha^2}{e^{m\alpha^2/\sigma^2}-1}\\
&=&\frac{1}{e^{m/\sigma^2}-1}\label{hcr-gaussian2},
\end{eqnarray}
where the maximum is attained for $\alpha=\pm 1$.  A point worth mentioning is the role of the prior information. While~\eqref{cr-gaussian} drops linearly, ~\eqref{hcr-gaussian2} decreases exponentially  with respect to the number of observations. It is also interesting to note that~\eqref{hcr-gaussian} applies as well to the case in which the parameter is not restricted. We then have to deal with the maximization in~\eqref{hcr-gaussian} for variations in $\alpha$, where $\alpha$ may take any value (not necessarily integer) except $\alpha=0$. Since the right hand side of~\eqref{hcr-gaussian} is a decreasing function of $\alpha$, we let $\alpha\rightarrow 0$ and get~\eqref{cr-gaussian}.
\end{example}

\subsection{Performance Lower Bound}
In the support recovery problem, we know a priori that each entry of the support vector takes values from the restricted set $\bm{\Delta}=\{1,2,\dots, p\}$. Hence, the HCR bound can provide us with a lower bound on the performance of any unbiased estimator of the support set.

\begin{theorem}
\label{prop:CSHCR}
Assume $\bm{\hat{\bm{s}}}(\bm{y})$ to be an unbiased estimator of the support $\bm{s}$. The HCR lower bound on the variance of  $\bm{\hat{\bm{s}}}(\bm{y})$ is given by
\begin{equation}
\label{hcr-support}
\mbox{tr}[\mbox{cov}(\hat{\bm{s}})] \geq  \max_{i\in\{2,\cdots,N\}} \frac{\|\bm{s}-\bm{s}_i\|^2}{e^{\|\bm{x}-p_{\bm{s}_i}\bm{x}\|^2/\sigma^2}-1},
\end{equation}
in which $p_{\bm{s}_i}\bm{x}$ denotes the projection of $\bm{x}$ onto $\S{\bm{s}_i}$.
\end{theorem}

\begin{IEEEproof}
Since our observations are of the form $\by=\bPhi\btheta+\bepsilon $, the set of unknown parameters $\bdelta$ consists of the support vector $\bs(\btheta)=(n_1,n_2,\dots,n_k)$ and the corresponding coefficients $\btheta_{\bs}=(\theta_{n_1},\theta_{n_2},\dots,\theta_{n_k})$. We are only interested in estimating the support, hence, $\bdelta_1=\bs(\btheta)$ and $\bdelta_2=\btheta_{\bs}$. Then
\begin{equation*}
\frac{ \mathbb{P}^2(\bm{y};\bm{\delta}')}{\mathbb{P}(\bm{y};\bm{\delta})}=\prod_{i=1}^{m}\!\frac{1}{\sqrt{2\pi}\sigma} e^{-\frac{(y_i-2x'_i+x_i)^2-2(x'_i-x_i)^2}{2\sigma^2}},
\end{equation*}
where $\bx'=\bPhi\btheta'$. Upon integration we get
\begin{equation*}
 \int_{\mathbb{R}^m} \frac{\mathbb{P}^2(\bm{y};\bm{\delta}')}{\mathbb{P}(\bm{y};\bm{\delta})}d\bm{y}=e^{\frac{\|\bm{x}-\bm{x}'\|^2}{\sigma^2}}.
\end{equation*}
Using the HCR bound~\eqref{Hammersley}, we derive
\begin{equation}	
\label{hcr1}
\mbox{tr}[\mbox{cov}(\hat{\bm{s}})] \geq \sup_{\bdelta'\neq \bdelta}\frac{\|\bm{s}-\bm{s'}\|^2}{e^{\|\bm{x}-\bm{x}'\|^2/\sigma^2}-1}.
\end{equation}
If $\bx$ and $\bx'$ live in the same subspace, i.e., $\bs=\bs'$, the right hand side of~\eqref{hcr1} will be zero. Therefore, in order to find the supremum, we can restrict our attention to all the signals which do not live in the same subspace as $\bx$ does:
\begin{equation}
\label{HCR-CS}
\mbox{tr}[\mbox{cov}(\hat{\bm{s}})] \geq \sup_{\{\btheta': \bs(\btheta')\neq \bs(\btheta)\}}\frac{\|\bm{s}(\btheta)-\bm{s}(\btheta')\|^2}{e^{\|\bm{x}-\bm{x}'\|^2/\sigma^2}-1}.
\end{equation}
For each sequence $\bm{s'}$, the numerator of~\eqref{HCR-CS} is fixed (it is the $\ell_2$ distance between the supports and does not depend on the coefficients) while the denominator is minimized by setting $\bm{x}' = p_{\bm{s'}}\bm{x}$. This leads to~\eqref{hcr-support}.
\end{IEEEproof}

\begin{corollary}
\label{corollary_HCR}
For any support vector $\bs_i \neq \bs$, we have
\begin{equation*}
\mbox{tr}[\mbox{cov}(\hat{\bm{s}})] \geq  \frac{\|\bm{s}-\bm{s}_i\|^2}{e^{\|\bm{x}-p_{\bm{s}_i}\bm{x}\|^2/\sigma^2}-1}.
\end{equation*}
\end{corollary}

In the following, we see how Theorem~\ref{prop:CSHCR} helps us to find the lower bound on the number of measurements for reliable $\ell_2$-norm support recovery.

% Necessary Conditions
\subsection{Necessary Conditions}
\label{subsec:necessary}
Using the HCR bound, Theorem~\ref{prop:CSHCR} provides a lower bound on the performance of any unbiased estimator for the $\ell_2$-norm support recovery problem. In words,  the $\ell_2$-norm support recovery is unreliable if the right hand side of~\eqref{hcr-support} is bounded away from zero, which yields to a lower bound on the minimum number of measurements. However, finding the maximum in~\eqref{hcr-support} requires a search through an exponential number of subspaces. Instead, as Corollary~\ref{corollary_HCR} suggests, any subspace different from the true one will provide us with a lower bound.  In the following, we show how this result will lead to necessary conditions for random Gaussian measurement matrices.

\begin{theorem}
\label{necessary}
Let the measurement matrix $\bPhi\in\R^{m\times p}$ be drawn with i.i.d. elements from a standard Gaussian distribution $\calN\left(0,1\right)$. The $\ell_2$-norm support recovery over the signal class $C(\theta_{\min})$ is unreliable if
\begin{equation*}
m < \max\left\{k, \frac{\sigma^2\log (p-k)}{\thetam^2}\right\}.
\end{equation*}
\end{theorem}

\begin{IEEEproof}
The $\ell_2$-norm support recovery is reliable if~\eqref{asym} holds for any $\btheta\in C(\theta_{\min})$. Consider a $\btheta$ with $\bs(\btheta)=(1,2,\dots,k)$ which takes $\thetam$ as its last non-zero entry, i.e., $\theta_k=\thetam$.
From Corollary~\ref{corollary_HCR},   we have
\begin{equation}\label{trcov}
\tr[\cov(\hat{\bs})] \geq  \frac{\|\bs-\bs'\|^2}{e^{\|\bx-\bx'\|^2/\sigma^2}-1},
\end{equation}
for any $\bx' = \bPhi \btheta' \in \bPhi_{\bs'}$. In particular, let $\btheta'$ have the support $\bs(\btheta')=(1,2,\dots,k-1,p)$ with coefficients equal to those of $\btheta$ in the first $k-1$ positions and $\theta'_p=\thetam$. We show that if $m$ does not satisfy the condition of the theorem, then the RHS of \eqref{trcov} will be bounded away from zero for this specific $\btheta'$, and therefore the estimation is unreliable.

Note that $\|\bs-\bs'\|^2 = (p-k)^2$, and $\bx-\bx'=\bPhi(\btheta-\btheta')$.
%\begin{equation*}
%\bx-\bx'=\bPhi(\btheta-\btheta').
%\end{equation*}
This implies that
\begin{equation*}
\frac{\|\bx-\bx'\|^2}{\sigma^2}= \frac{\thetam^2}{\sigma^2} \|\bPhi_{k} - \bPhi_{p}\|^2  = \frac{2\thetam^2}{\sigma^2}Z,
\end{equation*}
where $Z\sim \chi^2(m)$ has a chi-square distribution with $m$ degrees of freedom. It is known that a central chi-square random variable with $m$ degrees of freedom satisfies
\begin{equation}\label{tailz}
\Pr \left(Z-m\geq 2\sqrt{mt}\right)\leq e^{-t},
\end{equation}
for all $t\geq 0$~\cite{LaurentM98}. Assume that
\begin{equation}\label{condition}
m<(1-C)\frac{\sigma^2\log(p-k)}{\thetam^2},
\end{equation}
for some constant $C>0$, and evaluate \eqref{tailz} for $t=\left(\frac{\sigma^2\log(p-k)}{\thetam^2}-m\right)^2/4m$. This leads to
\begin{equation}\label{inter2}
\Pr\left(Z \geq \frac{\sigma^2\log(p-k)}{\thetam^2} \right) \leq \exp \left[\frac{-\left(\frac{\sigma^2\log(p-k)}{\thetam^2}-m\right)^2}{4m}\right].
\end{equation}
Note that the RHS of \eqref{inter2} converges to zero, as $p$ grows. Therefore,
\begin{align*}
\Pr\left(\frac{\|\bx-\bx'\|^2}{\sigma^2} < \log\left(p-k\right)^2 \right)&=\Pr\left(Z <\frac{\sigma^2\log(p-k)}{\thetam^2} \right) \\ \nonumber
&\rightarrow 1,
\end{align*}
which shows that the RHS of \eqref{trcov} is bounded away from zero with high probability, and therefore, the estimation error does not vanish asymptotically.

\end{IEEEproof}

Table~\ref{regime1} shows the necessary conditions for different scalings of $k$ and $\theta_{\min}$ as a function of $p$.

Up to this point, we have discussed the HCR bound and its application in finding necessary conditions on the number of measurements for reliable $\ell_2$-norm support recovery for Gaussian measurement matrices. In the following, we find conditions under which the HCR bound is achievable and as a result, find the sufficient number of measurements for reliable $\ell_2$-norm support recovery.

% Achievability
\section{Achievability of the HCR Bound}
 \label{sec:ml}

We now analyze the performance of the maximum likelihood estimator (MLE) for the $\ell_2$-norm support recovery and find conditions under which it becomes unbiased and in addition, its performance moves towards that of the HCR bound. We then apply this result to derive a sufficient number of measurements for the standard Gaussian measurement matrices.

\subsection{MLE performance}
Provided that any $2k$ columns of the measurement matrix $\bm{\Phi}$ are linearly independent, the noiseless measurement vector $\bm{x}=\bm{\Phi}\bm{\theta}$ belongs to one and \emph{only} one of the $N$ possible subspaces. Since the noise $\epsilon\in\mathbb{R}^{m}$ is i.i.d. Gaussian, MLE selects
the subspace closest to the observed vector $\bm{y}\in\mathbb{R}^{m}$. More precisely,
\begin{equation*}
\hat{\bm{s}}_{\text{\tiny{ML}}}=\underset{\bm{s}:|\bm{s}|=k}{\operatorname{argmin}}\:\:\|\bm{y}- p_{\bm{s}}\bm{y}\|.\label{eq:ml}
\end{equation*}
Now consider another subspace $\S{\bs'}$ of dimension $k$ where $\bm{s}\neq \bm{s}'$. Clearly an error happens when MLE
selects the support $\bm{s}'$ in place of the true support $\bm{s}$. Let $\Pr_{\tiny{\mbox{ML}}}(\bm{s}')$ denote the probability that MLE outputs the support vector $\bm{s}'$ instead of $\bm{s}$, among all possible support vectors.
\begin{lemma}
\label{ML-lemma}
Let $\bm{y} = \bm{x} + \bm{\epsilon}$, where $\bm{x}=\bm{\Phi}\bm{\theta}\in\S{\bs}$, $\bm{\epsilon}\sim \mathcal{N}(0,\sigma^{2}\bm{I})$ and $\bm{s}'$ be a support set different from $\bm{s}$. Then
\begin{equation*}
\Pr_{\tiny{\mbox{ML}}}(\bm{s}') < \Pr \left(\|\bm{\epsilon}\| \geq \frac{\|\bm{x}-p_{\bm{s}'}\bm{x}\|}{2}\right).
\end{equation*}
\end{lemma}
\begin{IEEEproof}
See Appendix~\ref{sec:appendixA}.
\end{IEEEproof}

Let the minimum distance between $\bm{x}$ and its projections onto other subspaces be
\begin{equation*}
d_{\text{min}}\triangleq \displaystyle\min_{\bm{s}': \bm{s}' \neq \bm{s}} \|\bm{x}-p_{\bm{s}'}\bm{x}\|,
\end{equation*}
 and the \emph{distinguishability factor} $\beta$ be defined as
\begin{equation*}
\beta= d_{\text{min}}^2/4m\sigma^2.
\end{equation*}

\begin{lemma}
\label{ML:lemma}
Let $\bm{y} = \bm{x} + \bm{\epsilon}$, where $\bm{x}=\bm{\Phi}\bm{\theta}\in \S{\bs}$ and $\bm{\epsilon}\sim \mathcal{N}(0,\sigma^{2}\bm{I})$. Moreover, assume that the number of measurements $m$ is an even integer, and $\beta>1$. Then, the probability that MLE makes an error in choosing $\bm{s}$ is upper bounded by
\begin{equation*}
\label{ML-Reg}
\Pr_{\tiny{\mbox{ML}}}(\mbox{err}) < \frac{m}{2}\: c(\beta)^{-\beta m},
\end{equation*}
where $c(\beta)= e^{(\beta-1)/2\beta}/\beta^{1/2\beta}> 1$ and $c(\beta)\longrightarrow \sqrt{e}$ as $\beta$ grows.
\end{lemma}

\begin{IEEEproof}
See Appendix~\ref{sec:appendixB}.
\end{IEEEproof}

%\begin{lemma}\label{ML:lemma}
%Let the number of measurements $m$ be an even integer. Then
%\begin{equation*}
%\label{ML-Reg}
%\Pr_{\tiny{\mbox{ML}}}(\bm{s}') < e^{-r/2}\displaystyle\sum_{t=0}^{m/2-1} \frac{(r/2)^t}{t!},
%\end{equation*}
%where $r = \frac{\|\bm{x}-p_{\bm{s}'}\bm{x}\|^2}{4\sigma^2}$.
%\end{lemma}
%\begin{IEEEproof}
%See Appendix~\ref{sec:appendixB}.
%\end{IEEEproof}
%\begin{lemma}
%\label{ML-thm}
%Let $r=\alpha m$ for some constant $\alpha > 1$. Then we have
%\begin{equation*}
%\label{ML-Asymp}
%\Pr_{\tiny{\mbox{ML}}}(\bm{s}') < \frac{r}{2\alpha} \:c(\alpha)^{-r},
%\end{equation*}
%in which $c(\alpha)= e^{(\alpha-1)/2\alpha}/\alpha^{1/2\alpha}> 1$ and $c(\alpha)\longrightarrow \sqrt{e}$ as $\alpha$ grows.
%\end{lemma}
%\begin{IEEEproof}
%See Appendix~\ref{sec:appendixC}.
%\end{IEEEproof}

Based on Lemma~\ref{ML:lemma}, the probability of error of MLE is related to the minimum distance between $\bm{x}$ and its projections onto the other subspaces. In the following theorem, we provide a bound on the performance of MLE.

\begin{theorem}
\label{ML:performance}
Let $\beta > 1$ and $m \geq (1+\varepsilon) \log{(p)}/(\beta\log{c(\beta)})$ for some fixed $\varepsilon > 0$. Then, MLE is asymptotically unbiased as $p\rightarrow \infty$, namely,
\begin{equation*}
\lim_{p \rightarrow \infty}\mathbb{E}(\bm{\hat{s}})=\bm{{s}}.
\end{equation*}
Moreover, its performance is bounded by
\begin{equation}
\label{covar-ML}
\mbox{\tr}[\underset{\tiny{\mbox{ML}}}{\mbox{\cov}}(\hat{\bm{s}})]< \frac{kmp^2}{2}  \:c(\beta)^{-\beta m},
\end{equation}
in which $c(\beta)= e^{(\beta-1)/2\beta}/\beta^{1/2\beta}> 1$ and $c(\beta)\longrightarrow \sqrt{e}$ as $\beta$ grows.
\end{theorem}
\begin{IEEEproof}
Let $\bm{\hat{s}}=(\hat{n}_1,\hat{n}_2,\dots,\hat{n}_k)$ be the ML estimate for the true support set $\bm{s} = (n_1,n_2,\dots,n_k)$. Then
\begin{eqnarray*}
\mathbb{E}(\bm{\hat{s}})&=& \sum_{i=1}^N \bm{s}_i\Pr_{\mbox{\tiny {ML}}}(\bm{s}_i)\\\nonumber
&=&\bm{{s}}\Pr_{\mbox{\tiny {ML}}}(\bm{{s}})+\sum_{\bm{s}_i\neq \bm{{s}}} \bm{s}_i\Pr_{\mbox{\tiny {ML}}}(\bm{s}_i).
\end{eqnarray*}
Since $\sum_{\bm{s}_i\neq \bm{{s}}}\Pr_{\mbox{\tiny{ML}}}(\bm{s}_i)=\Pr_{\tiny{\mbox{ML}}}(\mbox{err})$ and $1\leq\hat{n}_i\leq p$, we have
\begin{equation}
\label{eq:unbiaseq}
\sum_{\bm{s}_i\neq \bm{{s}}} \bm{s}_i\Pr_{\mbox{\tiny {ML}}}(\bm{s}_i)\leq (p,p,\dots,p)\Pr_{\tiny{\mbox{ML}}}(\mbox{err}).
\end{equation}
in which $\Pr_{\tiny{\mbox{ML}}}(\mbox{err})$ denotes the probability that MLE makes an error. Combining~\eqref{eq:unbiaseq} and Lemma~\ref{ML:lemma},  we get
\begin{equation*}
 \lim_{p \rightarrow \infty}\sum_{\bm{s}_i\neq \bm{{s}}} \bm{s}_i\Pr_{\mbox{\tiny {ML}}}(\bm{s}_i)\leq\lim_{p \rightarrow \infty} (p,p,\dots,p)\frac{m}{2} \:c(\beta)^{-\beta m}\overset{(a)}{=}\bm{0},
\end{equation*}
where in $(a)$, we used $m \geq (1+\varepsilon)\log{(p)}/(\beta\log{c(\beta)})$. Obviously, $ \Pr_{\mbox{\tiny {ML}}}(\bm{{s}})\rightarrow 1$ as $p\rightarrow \infty$. Hence $\lim_{p \rightarrow \infty}\mathbb{E}(\bm{\hat{s}})=\bm{{s}}$. For the second part, we need to compute the asymptotic behavior of $\mbox{\tr}[\underset{\tiny{\mbox{ML}}}{\mbox{\cov}}(\hat{\bm{s}})] $ as $p \rightarrow \infty$. By definition
\begin{equation*}
\mbox{\tr}[\underset{\tiny{\mbox{ML}}}{\mbox{\cov}}(\hat{\bm{s}})] = \mathbb{E}(\|\hat{\bm{s}} - \mathbb{E}(\hat{\bm{s}})\|^2).
\end{equation*}

Now, as $p \rightarrow \infty$ we can write
\begin{align*}
\mbox{\tr}[\underset{\tiny{\mbox{ML}}}{\mbox{\cov}}(\hat{\bm{s}})] &= \displaystyle\sum_{\bm{s}_i}  \Pr_{\tiny{\mbox{ML}}}( \bm{s}_i) \|\bm{s}_i - \mathbb{E}(\bm{\hat{s}})\|^2 \\ \nonumber
&\overset{(a)}{<} kp^2\displaystyle\sum_{\bm{s}_i\neq \bm{{s}}}  \Pr_{\tiny{\mbox{ML}}}( \bm{s}_i)  \\ \nonumber
 &\overset{(b)}{<} \frac{kmp^2 }{2} \:c(\beta)^{-\beta m},
\end{align*}
where in $(a)$ we used the fact that  $\|\bm{s}_i - \mathbb{E}(\bm{\hat{s}})\|^2$ is bounded by $kp^2$ and for $(b)$ we used Lemma~\ref{ML:lemma}.
\end{IEEEproof}

By Theorem~\ref{ML:performance}, MLE is asymptotically unbiased and therefore, its estimation error is lower bounded by the HCR bound. Moreover, the MLE performance upper bound in~\eqref{covar-ML} has only a 9 dB gap in the denominator compared to the HCR lower bound in~\eqref{hcr-support}. Therefore, such asymptotic behavior of MLE shows the achievability of the HCR bound, under the mentioned conditions.

As we observe, our results do not depend on any specific measurement matrix. In the following, we see how these results lead us to find the sufficient number of measurements for reliable $\ell_2$-norm support recovery when the Gaussian measurement ensemble is used.

% Simulations
\subsection{Sufficient Conditions}
\label{subsec:suff}
Theorem~\ref{ML:performance} provides us with a bound on the performance of the MLE. For reliable $\ell_2$-norm support recovery, the right hand side of (\ref{covar-ML}) should go to zero as $p \rightarrow\infty$. To that end,  as required by Theorem~\ref{ML:performance}, one should make sure that first, $\beta$ is bounded away from one which is a property of the underlying measurement matrix and second, that the number of measurements is at least of the order of $\log p$. Note that these conditions also imply that MLE is asymptotically unbiased and therefore, its performance is bounded by the HCR bound.

In the following, we study the above two conditions for random Gaussian measurement matrices, which will provide us with the sufficient number of measurements for reliable $\ell_2$-norm support recovery.
\begin{theorem}
\label{suff1}
Let the measurement matrix $\bPhi$ be drawn with i.i.d. elements from the standard Gaussian distribution $\mathcal{N}(0,1)$. If the minimum coefficient value of the signal satisfies $\frac{\theta_{\min}^2}{\sigma^2} > c$ for a constant $c$, then $m=\Theta(k\log\frac{p-k}{k})$ measurements suffice to ensure  reliable $\ell_2$-norm support recovery.
\end{theorem}
\begin{IEEEproof}
To ensure that $\beta > 1$, we need to find the scaling for which
\begin{equation}
\label{eq:maintoshow}
  \Pr\left(\min_{\bm{s}': \bm{s}' \neq \bm{s}} \|\bm{x}-p_{\bm{s}'}\bm{x}\| > 4m\sigma^2\right) \rightarrow 1, \quad \quad
\end{equation}
where $\bx = \bPhi \btheta$ and $\bs'$ goes through all support vectors different from $\bs = \bs_1$ (i.e., from $\bs_2$ to $\bs_N)$. We have,
\begin{equation*}
  \|\bx-P_{\bs'}\bx\|^2 = \|P_{\bs'}^\bot \bPhi \btheta\|^2.
\end{equation*}
Since the projection operator $P_{\bs'}^\bot $ cancels out any vector which lives in the subspace $\S{\bs'}$, we can write
\begin{equation*}
\|P_{\bs'}^\bot \bPhi \btheta\|^2 = \|P_{\bs'}^\bot\bPhi_{\bs/\bs'}\btheta_{\bs/\bs'}\|^2,
\end{equation*}
where $\bs/\bs'$ denotes the elements of $\bs$ which do not belong to $\bs'$. Now since
\begin{equation*}
 \frac{\|\bPhi_{\bs/\bs'}\btheta_{\bs/\bs'}\|^2}{\|\btheta_{\bs/\bs'}\|^2} \sim \chi^2(m),
\end{equation*}
and the range of the orthogonal projector $P_{\bs'}^\bot$ is of dimension $m-k$, we get
\begin{equation}
\label{eq:chi-square}
 X_{\bs,\bs'} \triangleq \frac{\|P_{\bs'}^\bot\bPhi_{\bs/\bs'}\btheta_{\bs/\bs'}\|^2}{\|\btheta_{\bs/\bs'}\|^2} \sim \chi^2(m-k).
\end{equation}

Let $\mathcal{A}_j$ denote the event $\{\bx : \|\bx-P_{\bs_j}\bx\|^2 > 4m\sigma^2\}$. Then,

\begin{align*}
  \Pr\left(\min_{\bm{s}': \bm{s}' \neq \bm{s}} \|\bx-P_{\bs'}\bx\|^2 > 4m\sigma^2\right) &= \Pr\left(\overset{N}{\underset{j=2}{\bigcap}} \: \mathcal{A}_j\right) \\
  &= \Pr\left(\left[\overset{N}{\underset{j=2}{\bigcup}} \mathcal{A}_j^c\right]^c\right)\\
  &\overset{(a)}{\geq} 1-\sum_{j=2}^{N} \Pr\left(\mathcal{A}_j^c\right),
\end{align*}
where in $(a)$ we used the union bound. In order to satisfy~\eqref{eq:maintoshow}, we seek conditions under which  the sum $\sum_{j=2}^{N} \Pr(\mathcal{A}_j^c)$ tends to zero. Each individual term in this sum can be written as
\begin{equation}
\label{eq:Ajc}
\Pr\left(\mathcal{A}_j^c\right) = \Pr\left( X_{\bs,\bs_j} \leq \frac{4m\sigma^2}{\|\btheta_{\bs/\bs_j}\|^2}\right).
\end{equation}
Since $X_{\bs,\bs_j} \sim \chi^2(m-k)$ (see~\eqref{eq:chi-square}), we can apply the following large deviation bound for the centralized $\chi^2$ distributions~\cite{LaurentM98}
\begin{equation}
\Pr\left(X_{\bs,\bs_j}-(m-k) \leq -2\sqrt{(m-k)x_j}\right) \leq e^{-x_j},
\label{eq:cond-xj}
\end{equation}
which is valid for all $x_j \geq 0$. Now, define
\begin{equation}
  x_j = \frac{(\frac{m-k}{2}-\frac{2m\sigma^2}{\|\btheta_{\bs/\bs_j}\|^2})^2}{m-k},
\label{eq:def-xj}
\end{equation}
and assume $\theta_{\text{min}}^2/\sigma^2>8$.
%\begin{equation*}
%  \frac{\theta_{\text{min}}^2}{\sigma^2} > 8.
%\end{equation*}
Hence, due to the fact that $2k < m$, we have
\begin{align}
\frac{2m\sigma^2}{\|\btheta_{\bs/\bs_j}\|^2}<  \frac{2m\sigma^2}{\theta_{\text{min}}^2} < \frac{1}{4} m < \frac{m-k}{2}.
\label{eq:cond1}
\end{align}
Therefore, by evaluating \eqref{eq:cond-xj} for $x_j$ in \eqref{eq:def-xj} and using~\eqref{eq:cond1} , we have

%and under the condition that
%\begin{equation}
%\label{eq:cond}
%  m-k > \frac{4m\sigma^2}{\|\btheta_{\bs/\bs_j}\|^2},
%\end{equation}
%the probability in~\eqref{eq:Ajc} can be upper bounded as

\begin{equation*}
\Pr\left(\mathcal{A}_j^c\right)  \leq \exp{\left(-\left(\frac{\sqrt{m-k}}{2}-\frac{2m\sigma^2}{\|\btheta_{\bs/\bs_j}\|^2\sqrt{m-k}}\right)^2\right)}.
\end{equation*}

%Note that since $\|\btheta_{\bs/\bs_j}\| \geq \theta_{\text{min}}$, the condition given in~\eqref{eq:cond} is satisfied (irrespective of the subspace $\bs_j$) provided that
%\begin{equation}
%\label{eq:cond1}
%  m-k > \frac{4m\sigma^2}{\theta_{\text{min}}^2}.
%\end{equation}
%Due to the fact that $2k < m$,~\eqref{eq:cond1} is satisfied if
%\begin{equation*}
%  \frac{\theta_{\text{min}}^2}{\sigma^2} > 8.
%\end{equation*}
Let $\ell_j = |\bs/\bs_j|$ be the number of indices in $\bs$ not present in $\bs_j$. Then
\begin{equation}
\label{eq:lj}
\|\btheta_{\bs/\bs_j}\|^2 \geq \ell_j \:\theta_{\text{min}}^2.
\end{equation}
Let the symbols $\bigtriangleup_j$ and $\bigtriangledown_{\ell_j}$ be defined as
\begin{eqnarray*}
  \bigtriangleup_j &=& \frac{\sqrt{m-k}}{2}-\frac{2m\sigma^2}{\|\theta_{\bs/\bs_j}\|^2\sqrt{m-k}}\\
\bigtriangledown_{\ell_j} &=& \frac{\sqrt{m-k}}{2}-\frac{2m\sigma^2}{\ell_j\theta_{\text{min}}^2\sqrt{m-k}}.
\end{eqnarray*}
Then, $\theta_{\text{min}}^2/\sigma^2>8$ and~\eqref{eq:lj} implies
\begin{equation*}
  \bigtriangleup_j \geq \bigtriangledown_{\ell_j} > 0.
\end{equation*}
%where the symbols $\bigtriangleup_j$ and $\bigtriangledown_{\ell_j}$ are defined as
%\begin{eqnarray*}
%  \bigtriangleup_j &=& \frac{\sqrt{m-k}}{2}-\frac{2m\sigma^2}{\|\theta_{\bs/\bs_j}\|^2\sqrt{m-k}}\\
%\bigtriangledown_{\ell_j} &=& \frac{\sqrt{m-k}}{2}-\frac{2m\sigma^2}{\ell_j\theta_{\text{min}}^2\sqrt{m-k}}.
%\end{eqnarray*}
and therefore,
\begin{equation}
\label{eq:exp-delta}
  \exp{(-\bigtriangleup_j^2)} \leq \exp{(-\bigtriangledown_{\ell_j}^2)}.
\end{equation}
Combining~\eqref{eq:Ajc} and~\eqref{eq:exp-delta} and taking summation over all possible error events, we get
\begin{align*}
  \sum_{j=2}^{N} \Pr\left(\mathcal{A}_j^c\right) &\leq \sum_{j=2}^{N} \exp{(-\bigtriangleup_j^2)} \\
  &\leq \sum_{j=2}^{N} \exp{(-\bigtriangledown_{\ell_j}^2)} \\
&\leq \sum_{\ell=1}^{k} {k \choose \ell}{p-k \choose \ell} \exp{(-\bigtriangledown_\ell^2)} \\
  &\leq k \max_{1 \leq \ell \leq k}\left\{{k \choose \ell}{p-k \choose \ell} \exp{(-\bigtriangledown_\ell^2)}\right\}.
\end{align*}
As we mentioned earlier, the sum $\sum_{j=2}^{N} \Pr(\mathcal{A}_j^c)$ should tend to zero as the dimension $p$ grows. This will hold if
\begin{equation}
\label{eq:infty}
  \lim_{p \rightarrow \infty}\max_{1 \leq \ell \leq k}\left\{ \log{k} + \log{k \choose \ell} + \log{p-k \choose \ell}-\bigtriangledown_\ell^2\right\} \rightarrow -\infty.
\end{equation}
Without loss of generality, we assume that $\sigma^2 = 1$. Let us define
\begin{equation*}
 \alpha_\ell \triangleq \frac{\bigtriangledown_\ell^2}{m-k}.
\end{equation*}
Applying~\eqref{eq:cond1}, it is easy to show that
\begin{equation*}
\alpha_{\ell} = \left(\frac{1}{2} - \frac{2m}{\ell \theta_{\text{min}}^2 (m-k)}\right)^2 \leq \frac{1}{4}.
\end{equation*}
Therefore using Stirling's approximation,~\eqref{eq:infty} is satisfied asymptotically if
\begin{equation}
  \label{eq:msuff}
  m > k + \max_{1 \leq \ell \leq k}\left\{ \log{k} + \ell\log{\frac{k}{\ell}+\ell\log{\frac{(p-k)}{\ell}}}\right\}.
\end{equation}
To find the maximum in~\eqref{eq:msuff}, we consider separately the linear and sub-linear regimes.
\begin{enumerate}
  \item $\ell = \Theta(k)$:\\
  We have
  \begin{equation*}
    m > c_1\log{k}+c_2k+c_3k\log{\frac{p-k}{k}},
  \end{equation*}
  for some constants $c_1, c_2, c_3$ greater than zero. Since $k\log{\frac{p-k}{k}}$ dominates the other terms asymptotically, we should have
  \begin{equation*}
    m = \Theta(k\log{\frac{p-k}{k}}).
  \end{equation*}
  \item $\ell = o(k)$:\\
  In this regime we have
  \begin{equation*}
     \ell \log{\frac{k}{\ell}} < k,
  \end{equation*}
  and
  \begin{equation*}
  \label{eq:proof1}
    \ell\log{\frac{p-k}{\ell}} < k\log{\frac{p-k}{k}}.
  \end{equation*}
  Therefore, the result of the linear regime covers the sub-linear regime.
\end{enumerate}
Thus, we showed that $m=\Theta(k\log\frac{p-k}{k})$ measurements is sufficient for perfect $\ell_2$-norm support recovery under the standard Gaussian measurement ensemble.
\end{IEEEproof}

Based on Theorem~\ref{suff1}, the sufficient number of measurements under different scalings for $k$ is given by
\begin{align*}
  k &= \Theta(p) \Longrightarrow m = \Theta(p), \\
  k &= o(p) \Longrightarrow m = \Theta(k\log{\frac{p}{k}}).
\end{align*}
The necessary and sufficient conditions in different regimes for the standard Gaussian measurement ensemble are shown in Table~\ref{regime1}.

\begin{remark}
The first row in Table~\ref{regime1} shows that one needs to take more measurements than the dimension of the signal in order to estimate the exact support set. This seems to be in contradiction with the concept of compressed sensing. One might think that this is an artifact of using this particular way of sampling. To show that this is not the case, let us assume that we have direct access to the noisy version of the input signal $\btheta$. This means that we use a square diagonal matrix $\bm{D}$ instead of a Gaussian one to sample the signal. In order to make the two scenarios comparable, we should make sure that the signal powers after the measurement are equal. To this end, we need to put a gain of $\sqrt{k}$ on the main diagonal.

Now consider two signals $\btheta_1$ and $\btheta_2$ which consist of $k$ nonzero entries with amplitudes $\theta_{\text{min}}$ and differ in only one position. The probability of error of MLE is given by

\begin{align*}
   \Pr_{\tiny{\mbox{ML}}}(\mbox{err}) &= \mathbb{Q}\left(\frac{\|\bm{D}\btheta_1-\bm{D}\btheta_2\|}{2\sigma}\right) \\
   &= \mathbb{Q}\left(\frac{\sqrt{2k\theta_{\text{min}}^2}}{2\sigma}\right),
\end{align*}
where $\mathbb{Q}(\cdot)$ is the tail probability of a standard Gaussian random variable. In the regime considered in the first row of  Table~\ref{regime1}, i.e., $\frac{\theta_{\min}^{2}}{\sigma^2}=\Theta\left(\frac{1}{k}\right)$ we obtain
\begin{equation}
  \Pr_{\tiny{\mbox{ML}}}(\mbox{err}) = \mathbb{Q}(\text{constant}) > 0.
\end{equation}
 Therefore, even if we use direct measurements, there is no hope to recover the exact support in this regime. In~\cite{Wainright07}, Wainwright showed that $\Theta(p\log{p})$ measurements is indeed sufficient.
\end{remark}

\begin{table}
\begin{center}  \begin{small}
\begin{tabular}{|c|c|c|}
\hline
 & Necessary  & Sufficient \tabularnewline
\hline
\hline
\begin{tabular}{c}
$k=\Theta(p)$\tabularnewline
$\theta_{\min}^{2}=\Theta\left(\frac{1}{k}\right)$\tabularnewline
\end{tabular} & $\Theta\left(p\log p^{}\right)$ & $\blacksquare$\tabularnewline
%\hline
%\begin{tabular}{c}
%$k=\Theta(p)$\tabularnewline
%$\theta_{\min}^{2}=\Theta\left(\frac{\log k}{k}\right)$\tabularnewline
%\end{tabular} &$\Theta\left(p\right)$ & $*$\tabularnewline
\hline
\begin{tabular}{c}
$k=\Theta(p)$\tabularnewline
$\theta_{\min}^{2}=\Theta\left(1\right)$\tabularnewline
\end{tabular} & $\Theta( p)$& $\Theta( p)$\tabularnewline
\hline
\begin{tabular}{c}
$k=o(p)$\tabularnewline
$\theta_{\min}^{2}=\Theta\left(\frac{1}{k}\right)$\tabularnewline
\end{tabular} & $\Theta\left(k\log(p-k)\right)$& $\blacksquare$\tabularnewline
%\hline
%\begin{tabular}{c}
%$k=o(p)$\tabularnewline
%$\theta_{\min}^{2}=\Theta\left(\frac{\log k}{k}\right)$\tabularnewline
%\end{tabular} & $\Theta\left(\frac{k\log(p-k)}{\log k}\right)$ & $*$\tabularnewline
\hline
\begin{tabular}{c}
$k=o(p)$\tabularnewline
$\theta_{\min}^{2}=\Theta\left(1\right)$\tabularnewline
\end{tabular} & $\max\{\Theta\left(k\right),\Theta\left(\log(p-k)\right)\}$ & $\Theta\left(k\log \frac{p}{k}\right)$\tabularnewline
\hline
\end{tabular}
\end{small}
\end{center}
\caption{Necessary and sufficient conditions on the number of measurements required for reliable $\ell_2$-norm support recovery under the standard Gaussian measurement ensemble ($\sigma^2 = 1$).}\label{regime1}
\end{table}

% Conclusions
\section{Conclusions}
\label{sec:Conclusions}
We considered the problem of recovering the support of a sparse vector from a set of noisy linear measurements from an estimation theoretic point of view. We set the error metric between the true and the estimated support sets as the $\ell_2$-norm of their differences. Then, we investigated the fundamental performance limit of any unbiased estimator of the support set using the Hammersley-Chapman-Robbins bound, where no specific assumption was made on the measurement matrix. This general bound led us to the necessary conditions on the number of measurements for successful support recovery, which we specifically evaluated for standard random Gaussian measurement ensembles. Then, we analyzed the performance of the maximum likelihood estimator and derived  conditions under which it becomes unbiased and achieves the Hammersley-Chapman-Robbins bound. Applying these conditions provided us with the sufficient number of measurements for  random Gaussian measurement ensembles.

% Acknowledgement
\section*{Acknowledgment}

The authors would like to thank Prof. T. Blu, Prof. M. J. Wainwright and Prof. R. \"{U}rbanke for their help and useful comments.

\appendices
\numberwithin{equation}{section}
\begin{section}{}
\label{sec:ProofLemmas}

\subsection{Proof of Lemma~\ref{ML-lemma}}
\label{sec:appendixA}

MLE chooses $\bm{s}'$ over $\bm{s}$ if and only if
\begin{equation*}
\label{ML}
\displaystyle\min_{\bm{t}' \in \bPhi_{\bm{s}'}} \|\bm{y}-\bm{t}'\| < \displaystyle\min_{\bm{t} \in \bPhi_{\bm{s}}} \|\bm{y}-\bm{t}\|.
\end{equation*}
Let us assume that
\begin{equation}
\label{eq:halfDist}
\|\bm{\epsilon}\| < \frac{\|\bm{x}-p_{\bm{s}'}\bm{x}\|}{2}.
\end{equation}
For any $\bm{t}' \in \bPhi_{\bm{s}'}$, we have
\begin{align*}
\|\bm{y}-\bm{t}'\|^2 &= \|\bm{x}-\bm{t}'+ \bm{\epsilon}\|^2 \\ \nonumber
%&= \|\bm{\epsilon}\|^2 + \|\bm{x} - \bm{t}'\|^2 + 2\|\bm{x} - \bm{t}'\|\|\bm{\epsilon}\|\cos{\theta} \\ \nonumber
&\geq \|\bm{\epsilon}\|^2 + \|\bm{x} - \bm{t}'\|^2  - 2\|\bm{x} - \bm{t}'\|\|\bm{\epsilon}\| \\ \nonumber
%&= \|\bm{\epsilon}\|^2 + 2\|\bm{x} - \bm{t}'\|\left(\frac{\|\bm{x} - \bm{t}'\|}{2}-\|\bm{\epsilon}\|\right) %\\\nonumber
&\overset{(a)}{>} \|\bm{\epsilon}\|^2 \\ \nonumber
&= \|\bm{y}-\bm{x}\|^2 \\ \nonumber
&\geq \displaystyle\min_{\bm{t} \in \bm{s}} \|\bm{y}-\bm{t}\|^2,
\end{align*}
where in $(a)$ we used~\eqref{eq:halfDist}. This implies that if $\|\bm{\epsilon}\| < \|\bm{x}-p_{\bm{s}'}\bm{x}\|/2$, MLE will not choose $\bm{s}'$ over $\bm{s}$. Since the probability that MLE selects $\bm{s}'$ among all possible support vectors is less than the probability that MLE chooses $\bm{s}'$ over $\bm{s}$, we get
\begin{equation*}
\Pr_{\tiny{\mbox{ML}}}(\bm{s}') < \Pr \left(\|\bm{\epsilon}\| \geq \frac{\|\bm{x}-p_{\bm{s}'}\bm{x}\|}{2}\right).\tag*{$\blacksquare$}
\end{equation*}

\subsection{Proof of Lemma~\ref{ML:lemma}}
\label{sec:appendixB}
From Lemma~\ref{ML-lemma} we know that if $\|\bm{\epsilon}\| < d_{\text{min}}/2$, MLE makes the correct choice. Therefore,
\begin{align*}
  \Pr_{\tiny{\mbox{ML}}}(\mbox{err}) &< \Pr \left(\|\bm{\epsilon}\| \geq d_{\text{min}}/2\right) \\ \nonumber
  &= 1-\Pr \left(\frac{\|\bm{\epsilon}\|^2}{\sigma^2}< r\right),
\end{align*}
where $r = \beta m$ and $\beta$ is the distinguishability factor. The random variable $\frac{\|\bm{\epsilon}\|^2}{\sigma^2}$ is distributed according to the chi-square distribution with $m$ degrees of freedom. By using the cumulative distribution function of the chi-square distribution, we obtain
\begin{equation}\label{gamma-ML}
\Pr_{\tiny{\mbox{ML}}}(\mbox{err}) < 1-\frac{\gamma(m/2,r/2)}{\Gamma(m/2)},
\end{equation}
where $\Gamma(m)$ is the Gamma function, and $\gamma(m,x)$ is the lower incomplete Gamma function. It is easy to show that for an even number $m$,
\begin{equation*}
\frac{\gamma(m/2,r/2)}{\Gamma(m/2)} = e^{-r/2} \displaystyle\sum_{t=\frac{m}{2}}^{\infty} \frac{(r/2)^t}{t!}.
\end{equation*}
Since by Taylor expansion $e^{r/2}=\sum_{t=0}^{\infty} \frac{(r/2)^t}{t!}$, we obtain
\begin{equation}\label{taylor}
\frac{\gamma(m/2,r/2)}{\Gamma(m/2)} = 1- e^{-r/2}\displaystyle\sum_{t=0}^{\frac{m}{2}-1} \frac{(r/2)^t}{t!}.
\end{equation}
Combining~\eqref{gamma-ML} and~\eqref{taylor}, we have
\begin{equation}
\label{lemma2:interm}
  \Pr_{\tiny{\mbox{ML}}}(\mbox{err}) < e^{-r/2}\displaystyle\sum_{t=0}^{m/2-1} \frac{(r/2)^t}{t!}.
\end{equation}
Note that for $t<\frac{r}{2} ,\: t \in \mathbb{N}$, the function $f(t) = \left(\frac{r}{2}\right)^{t}/t!$ is strictly increasing. Therefore,from~\eqref{lemma2:interm} we get

\begin{align*}
\Pr_{\tiny{\mbox{ML}}}(\mbox{err}) &<e^{-r/2}\displaystyle\sum_{t=0}^{\frac{m}{2}-1}\frac{(r/2)^t}{t!} \\ \nonumber
&\overset{(a)}{<}e^{-r/2}\frac{m}{2} \frac{(r/2)^{m/2}}{(m/2)!} \\ \nonumber
&\overset{(b)}{<} e^{-r/2} \frac{m}{2} \frac{(r/2)^{m/2}}{(m/2e)^{m/2}} \\ \nonumber
&= \frac{r}{2 \beta} \left(\frac{e^{(\beta-1)/2\beta}}{\beta^{1/2\beta}}\right)^{-r} \\  \nonumber
&= \frac{m}{2} c(\beta)^{-\beta m},
\end{align*}
where in $(a)$ we used $\frac{m}{2}<\frac{r}{2}$ and in $(b)$ we used the inequality $m! > (m/e)^m$. It can be easily verified that $c(\beta) > 1$ for $\beta > 1$ and  $c(\beta)\longrightarrow \sqrt{e}$ as $\beta$ grows.\[\tag*{$\blacksquare$}\]

\end{section}

% References
\bibliographystyle{ieeetr}
\bibliography{IEEEabrv,refs}
\end{document}